\documentclass[12pt,preprint]{aastex}

\newcommand{\psr}{PSR J0538+2817}

\shorttitle{Proper motion of \psr}
\shortauthors{Kramer et al.}

\slugcomment{Revised version of MS17462}

\begin{document}

\title{Proper motion, age and initial spin period of
PSR J0538+2817 in S147}

\author{M.~Kramer\altaffilmark{1},
        A.G. Lyne\altaffilmark{1},
        G.~Hobbs\altaffilmark{2,1},
        O.~L\"ohmer\altaffilmark{3},
        P.~Carr\altaffilmark{1}, 
        C.~Jordan\altaffilmark{1},
        A.~Wolszczan\altaffilmark{4}}
\altaffiltext{1}{University of Manchester, Jodrell Bank Observatory, Macclesfield, Cheshire SK11 9DL, UK}
\altaffiltext{2}{Australia Telescope National Facility, CSIRO, P.O.~Box~76, Epping, NSW~1710, Australia}
\altaffiltext{3}{Max Planck Institut f\"ur Radioastronomie, Auf dem H\"ugel 69, D-53121 Bonn, Germany}
\altaffiltext{4}{Department of Astronomy and Astrophysics, Pennsylvania State
University, University Park, PA 16802, USA}
\email{mkramer@jb.man.ac.uk}

\begin{abstract}
We present results of timing observations of the 143-ms pulsar
J0538+2817 that provide a proper motion measurement which clearly
associates the pulsar with the supernova remnant S147. We measure a
proper motion of 67$_{-22}^{+48}$ mas yr$^{-1}$,
implying a transverse
velocity of $v= 385^{+260}_{-130}$ km s$^{-1}$. We derive an age of
the pulsar and S147 of only $30\pm4$ kyr which is a factor 
of 20 times less
than the pulsar's characteristic age of $\tau_c = 620$ kyr.  This age
implies an initial spin period of $P_0=139$ ms, close to the present
pulse period and a factor of several larger than what is usually
inferred for birth periods. Implications for recent X-ray detections
of this pulsar are discussed.
\end{abstract}


\section{Introduction}
\label{intro}

The determination of the birth properties of pulsars is of crucial
importance in understanding both the physics of core-collapse
supernovae as well as the population and evolution of a radio
pulsar. The initial spin period of pulsars, $P_0$, is particularly
difficult to measure as it requires the knowledge of the pulsar's
age, $\tau$, and its spin-down behaviour. It is usually assumed that
the observed evolution of the spin-frequency $\nu = 1/P$ can be
described by a power-law $\dot{\nu} \propto -\nu^n $, where $n$ is
the so-called {\em braking index}.  The initial spin period can then
be calculated from
\begin{equation}
\label{eqn:p0}
P_0 = P \left[ 1 - \frac{n-1}{2}\;\frac{\tau}{\tau_c} 
\right]^{\frac{1}{n-1}}
\end{equation}
where $\tau_c=P/2\dot{P}$ is the {\em characteristic age} of the
pulsar. The characteristic age is a good estimator for the true
age of the pulsar, $\tau_c\approx \tau$, 
under the assumptions that $P_0\ll P$ 
and that the spin-down is due to magnetic braking
for which $n=3$.

Since the use of the characteristic age, rather than the true age, can
lead to considerable errors, it is desirable to have an independent
age measurement. The age of a supernova remnant (SNR) that originated
in the same explosion as the pulsar, can serve as such an estimator
but the only pulsar for which the age of an associated SNR is clearly
known, is the Crab pulsar. From the observation of the explosion in
A.D.1054 and a measured braking index of $n=2.51(1)$, the initial spin
period is computed to be $P_0=$19 ms (Lyne et al.~1993).\nocite{lps93}
It may also be possible to measure a proper motion of a pulsar associated
with a SNR.  If this transverse motion is directed away from the
center of the SNR, then this is strong evidence that the pulsar is
genuinely associated with the remnant.  It also allows one to
determine the age of both the pulsar and the SNR by comparing the
present offset from the center with its speed.  As SNRs typically fade
away after $\sim100,000$ yr, pulsars genuinely associated with SNRs
are necessarily young. Such pulsars, however, often show rotational
instabilities in the form of glitches and/or timing noise (e.g.~Lyne
et al.~1995)\nocite{lps95} which makes the measurement of proper
motion via timing observations usually a difficult
task. Interferometric measurements offer a solution but so far it has
only been possible for PSR B1951+32 in CTB 80 for which an initial
spin period of $P_0 =27(6)$ ms was derived (Migliazzo et
al.~2002\nocite{mgb+02}).

In this work we present a timing proper motion measurement for the
143-ms pulsar J0538+2817 which was found within the boundaries of
G180.0$-$1.7, also called S147 (Anderson et al.~1996\nocite{acj+96}).
The SNR S147 has a prominent shell structure with a radius of $\theta
= 83(3)$ arcmin (Sofue et al.~1980). With an estimated age of $\sim
100$ kyr (e.g.~Kundu et al.~1980\nocite{kafh80}) it is considered to
be one of the oldest well-defined SNRs in the Galaxy, although other
authors derived much younger ages (e.g.~20 kyr, Sofue et
al.~1980). Therefore, S147 has been studied rather extensively at
radio frequencies (e.g.~F\"urst et al.~1982\nocite{frb+82}).  Very
recently, X-ray observations with CHANDRA have revealed a structure
interpreted as a pulsar wind nebula (Romani \& Ng 2003\nocite{rn03})
while XMM-Newton observations revealed pulsed X-ray emission from the
surface (McGowan et al.~2003\nocite{mkz+03}).  In the following we
describe our timing observations and data
analysis which leads to a proper motion measurement which clearly
associates the pulsar with SNR S147. We hence obtain accurate
estimates of the age of the pulsar and derive its initial spin
period. The results are finally compared to the recent X-ray
observations.

\section{Observations and Data Analysis}

The observations were made with the 100-m Effelsberg radiotelescope
from April 1994 at 1410 MHz and with the 76-m Lovell telescope at
Jodrell Bank from March 1996 at 606 MHz and 1400 MHz.  At both
telescopes, two circularly polarized signals were mixed down to
intermediate frequencies, detected and incoherently de-dispersed in
hardware using filterbanks before sub-integrations of 15 sec
(Effelsberg) and 60 sec (Jodrell Bank) were written to disk for
off-line processing. All data were time-stamped with clock information
provided by local H-maser clocks which were later synchronized to UTC
by using signals from the Global Positioning System (GPS) satellites.
Details of the observing systems can be found in Anderson et
al.~(1996) and Hobbs et al.~(2003a).\nocite{hlk+03}

The data from both telescopes were first processed to a common time
resolution of 238.3 $\mu$s before being subjected to the same template
matching procedure which used identical templates to determine the pulse
times-of-arrival (TOAs). Since PSR J0538+2817 shows mode-changing,
exhibiting two distinct profiles which differ in the relative
height of the two prominent components (Anderson et al.~1996), we
applied a proven technique that uses two different, carefully created
templates for the given modes (see e.g.~Stairs et al.~2000).  
Any possible remaining effects due to the occurrence of
mixed-mode profiles are accounted for by adopting relatively large
minimum errors of $\Delta t\ge180\mu$s (Effelsberg) and $\Delta
t\ge130\mu$s (Jodrell Bank), respectively.  The success of this
procedure is notable by the fact that no arbitrary clock offsets
between the telescopes or profile modes were needed in the further
timing analysis. 

In a first step of the timing analysis using the DE200 planetary
ephemerides, the dispersion measure (DM) was determined by fitting for
DM and a simple spin-down model to data obtained with the Lovell
telescope at 606 and 1400 MHz over a small period of time. The DM value
was then held fixed for the subsequent analysis.  
Like many
other young pulsars, PSR J0538+2817 shows long-term timing noise
visible in the timing residuals. Hobbs et al.~(2003a)\nocite{hlk+03}
developed a new technique to remove such timing noise, allowing one
to reliably separate the signature due to proper motion.  Applying
this technique, the transverse proper motions have been determined for
more than 300 pulsars and excellent agreement is found for those
pulsar where interferometric measurements are available (Hobbs et
al.~2003a,b)\nocite{hlk03,hlk+03}. We applied this technique,
before fitting to a spin-down model which now included proper motion and a
second period derivative. Iterative tests were made to check
the robustness of the obtained solution which also studied the effects
of omitting and including different parts of the data in the analysis.
We do not believe that the non-zero second period derivative 
is due to magnetospheric braking, therefore we do not attempt to calculate
a braking index, $n$ (see Hobbs et al.~2003a).

As the pulsar lies close to the ecliptic with an ecliptic latitude,
$\beta$, of only $\beta=4^\circ.9$, 
position and proper motion measurements in the
latitudinal direction are necessarily much less accurate than those in
ecliptic longitude.  In order to minimize covariances between the
astrometric parameters, fits were made in ecliptic coordinates using
the software package {\tt TEMPO}\footnote{\tt
http://www.atnf.csiro.au/research/pulsar/timing/tempo}, resulting
in post-fit residuals shown in Fig.~1.  The final
spin and astrometric parameters are presented in
Table~\ref{tab:parms}. Quoted uncertainties are derived from twice the
formal {\tt TEMPO} error and standard Monte-Carlo  simulations.
Procedures for the latter are detailed in Lange et
al.~(2001)\nocite{lcw+01} and results obtained for proper motion are
shown in Fig.~\ref{fig:mc}.

\section{Results and Discussion}

As can be seen from Table~\ref{tab:parms} and Fig.~\ref{fig:mc}, we
have obtained a proper motion measurement for PSR J0538+2817. Whilst
the movement in ecliptic longitude is measured to high significance,
the uncertainty for the proper motion in latitude is much larger.
With a probability of 78\%, the proper motion is positive in
latitudinal direction, making the pulsar moving in the right quadrant
on the sky to be consistent with a movement away from the center of
S147.  We measure a position angle of P.A.$={311^\circ}^{+28}_{-56}$.  The
obtained proper motion is consistent with a comparison of the
interferometric position obtained by Anderson et al.~(1996)
and our present timing position.

Using the measured values, we can compute the pulsar's motion in the
past and compare it to the location of the center of the SNR.  Fitting
a circular shape 
to the radio contours of S147 over a frequency range from 430
MHz to 4750 MHz (Kundu et al.~1980\nocite{kafh80}, Sofue et 
al.~1980\nocite{sfh80}, Angerhofer \& Kundu 1981\nocite{ak81}, 
F\"urst et 
al.~1982\nocite{frb+82}, F\"urst \& Reich 1986\nocite{fr86})
we determine the SNR center to be at
$\lambda_{SNR}=85^\circ.57(1)$  and $\beta_{SNR}=4^\circ.44(1)$ which
agrees well with the center determined by Sofue et al.~(1980) at 4750
MHz only. We mark the central position on the 2.7-GHz map
obtained by F\"urst \& Reich~(1986) shown in Fig.~\ref{fig:map}. We
also mark the position of the pulsar 30 kyr ago as computed from the
measured proper motion ($\lambda_{30k}=85^\circ.57(3)$,
$\beta_{30k}=4^\circ.5(5)$). It is clear that the previous position of
the pulsar agrees very well with the center of the SNR, strongly
suggesting that the pulsar was born in the same explosion that created
S147, about 30,000 years ago.

Assuming that the pulsar was born at the center of S147, we use the
offset of the pulsar from this position, $\Delta \Theta =
2.18(4)\times 10^6 $ mas and the measured proper motion to determine a
true age of pulsar and remnant of $\tau= 33_{-9}^{+17}$ kyr. The
uncertainty in this age is dominated by the error in the proper motion
measurement in latitudinal direction.  In order to derive more
accurate estimate, we can use the offset and motion in longitudinal
direction only. This results in an age of $\tau_\lambda = 30\pm4$ kyr,
confirming that the kinematic age of the pulsar is dramatically smaller
than the characteristic age of $\tau_c= 618$ kyr. The only assumption
made in deriving this age is that the pulsar was born in the center of
the SNR. 

Anderson et al.~(1996) already discussed the possible
association of PSR J0538+2817 with S147 in detail and concluded that
an association is plausible. They based their arguments on the
proximity of the pulsar to the SNR center and the consistent distance
estimates for both pulsar and SNR. The distances estimated for S147
ranging from of 0.8 to 1.6 kpc are well consistent with the dispersion
measure distance of the pulsar, i.e. $d = 1.2$ kpc as 
derived from the NE2001 model (Cordes \& Lazio 2002\nocite{cl02}). 
At this distance, our proper motion measurement yields a
transverse speed of $v=385_{-130}^{+260}$ km s$^{-1}$ which is in
excellent agreement with mean observed velocity of pulsars (Lyne \&
Lorimer 1993).  Given the position of the pulsar within the SNR
boundaries and a normalized angular distance of only $\delta = \Delta
\Theta/ \Theta = 0.43(2)$ away from the center, and
its location in the Galactic anti-center region where we
find only a rather sparse population of known SNRs and pulsars, an
association seemed indeed very likely. This is now confirmed by the
pulsar's movement away from the center.

Further independent evidence is available that the characteristic age
is much larger than the true age of the pulsar.
Firstly, while some authors estimate a blast wave age of
S147 in a range from 80 kyr to 200 kyr (e.g.~Kundu et al.~1980), Sofue
et al.~(1980) estimated an age of only 20 kyr. Even though these age
estimates for SNRs depend also on the density of the ambient
interstellar medium and are known to be highly uncertain (e.g.~F\"urst
\& Reich 1986), all values are lower than $\tau_c$, 
and the latter estimate agrees indeed very well with our
findings. Given this young age of the SNR, the well defined shell
structure appears less surprising. If the explosion occurred in a
low-density, hot stellar wind cavity blown up by the progenitor star,
the expansion will not be describable by a Sedov-phase but will be
free until it reaches the cavity boundaries
(E.~F\"urst, private communication). At a distance of 1.2~kpc, the
observed SNR radius corresponds to $\sim 30$ pc, implying an
expansion velocity of 1000 km s$^{-1}$ in the free expansion
phase. Future CO observations may be able
to reveal the wind cavity.

Secondly, we have access to polarization information for the radio
emission of PSR J0538+2817 obtained by Mitra et al.~(2003), who
measured a rotation measure of RM=$-7\pm12$ rad
m$^{-2}$.  The pulsar exhibits an extremely high degree of
polarization of 92(2)\%. While this is not uncommon, it is
usually found in young sources (e.g.~Morris et al.~1981), 
again supporting a young age of the pulsar.

Finally, we can compare our results to recent X-rays observations.
Romani \& Ng (2003)\nocite{rn03} reported the discovery of a faint
nebula surrounding the pulsar.  They interpret this
nebula as an equatorial torus, supporting the association of pulsar
and S147. In their calculation they assumed a pulsar age of 100 to 200
kyr, but we can derive consistent results with our smaller derived age
of $\tau=30$ kyr with a somewhat smaller ISM density. The orientation
of their modelled torus with a symmetry axis located at position angle
$\Psi=154(6)^\circ$ (measured N through E) and $\xi=100(6)^\circ$
(into plane of the sky) is consistent with the position angle swing
measured for our polarization data. Applying a rotating vector model
(Radhakrishnan \& Cooke 1969)\nocite{rc69a}, we find a best fit for a
magnetic inclination of $\alpha \sim 95^\circ$ and an impact angle of
$\sigma\sim 2^\circ$ although the uncertainties are considerable. These
findings support Romani \& Ng's interesting conclusion derived from
the torus geometry that the velocity direction appears to be nearly
aligned with the rotation axis of the pulsar.  However, with a pulse
width of about $40^\circ$ and such a geometry, one may expect to
observe also emission from the opposite magnetic pole. This is not the
case, but the emission beam may not be circular and/or the emission
beam could be patchy (Lyne \& Manchester 1988). We also note that
the bilateral symmetry axis of S147 appears to be very nearly
parallel to both the symmetry axis of the torus and the proper motion
direction of the pulsar. While the remnant is somewhat asymmetric
about the orthogonal axis, the center of the explosion could in
principle deviate from the SNR center which is determined by
the sharp circular rim. However, it is reasonable to assume that the
explosion center is located on the symmetry axis between the
determined SNR center and the current pulsar position.  In such a
case, the age of the pulsar would be even younger, and our conclusions
would be essentially unaffected.

Equally intriguing is the comparison of our results with those by
McGowan et al.~(2003) \nocite{mkz+03} who detected pulsed X-ray
emission from PSR J0538+2817. Their data are well explained by
blackbody radiation from a heated polar cap. However, using the
dispersion measure distance the derived temperature is significantly
higher than predicted by standard cooling theories. This discrepancy
can be reduced by using atmospheric fits, but with a pure-H
non-magnetized atmosphere McGowan et al.'s result still falls above
the expected temperature. In their analysis McGowan et al.~used the
characteristic age of the pulsar, and the high temperature indeed
suggests that the true age of the pulsar is much smaller.  Using our
derived age of $\tau=30$ kyr, the temperature from the atmospheric fit
falls well below the standard cooling curve.  However, this fit also
produces a distance which is much smaller than the dispersion measure
estimate. A proper modelling of the observed X-ray spectra needs to
include the surface magnetic field of $B=7.3\times 10^{11}$ G, which
may change the results considerably (e.g. Pavlov et al.~2001, Zavlin
\& Pavlov 2002).\nocite{zp02}\nocite{pzs+01} This is important
since with such a low temperature one would be forced to
consider the presence of exotic cooling processes. Interestingly, a
similar conclusion has been reached recently by Slane et
al.~(2002)\nocite{shm02} for PSR J0205+6449 associated with 3C58 (but
see also Yakovlev et al.~2002\nocite{ykhg02}).

Pavlov et al.~(2002)\nocite{pzst02} studied CHANDRA data of the
neutron star 1E 1207.4$-$5209 which is likely to be associated with
SNR PKS 1209$-$51/52. They also find a characteristic age which is
much larger than that estimated for the SNR, suggesting a long birth
period of the neutron star.  Romani \& Ng already already pointed out
that the initial spin period of PSR J0538+2817 is likely to be large
and close to the present value.  Using our age estimate we derive an
initial spin period of $P_0 = 139.6$ ms ($n=3$), which is insensitive
to the actual choice of the braking index as the ratio of
$\tau/\tau_c=0.05$ is very small (e.g. $P_0 = 139.6 $ ms for $n=10$
and $P_0 =139.8$ ms for $n=0.5$).  It has been attempted to estimate
the initial spin period for seven radio pulsars (see Migliazzo et
al.~2002 and references therein) but apart from the results for the
Crab (based on the known age of the SNR and pulsar) and PSR B1951+32
in CTB 80 (based on kinematic age derived from proper motion), all
estimates rely on less certain ages estimated for associated SNRs. It
appears that the initial period of those pulsars is $P_0\le 60$ ms. An
exception is PSR J1124$-$5916 with $P_0\approx 90$ ms (Camilo et
al.~2002)\nocite{cmg+02}. All estimated birth periods are already
significantly larger than what is expected from core-collapse theory
of massive stars and it appears difficult to explain even spin periods
of a few tens milliseconds (see Heger et al.~2003\nocite{hwls03} for a
recent review). The estimated initial spin period for PSR J0538+2817
is much larger still, placing strong constraints on the origin of
birth kicks imparted on neutron stars as discussed by Romani \& Ng
(2003).

In summary, we have measured the proper motion of PSR J0538+2817 which
clearly associates the pulsar with the SNR S147. From the separation
of the pulsar from the SNR center we determine an age of $\tau
=30\pm4$ kyr, making the pulsar significantly younger than is
indicated by its characteristic age of $\tau_c=618$ kyr. This implies
a large initial spin period of $P_0=139$ ms. The implied possibility
of exotic cooling should be revisited after a magnetized atmosphere
has been fitted to the recent X-ray data.

\section*{Acknowledgements}

We thank Ernst F\"urst for useful disucssions and the 2.7-GHz map of
S147.  We also thank Werner Becker and Slava Zavlin for helpful
discussions and are grateful to an anonymous referee for his comments.


\begin{thebibliography}{}

\bibitem[Anderson {\rm et~al.}~{1996}]{acj+96}
Anderson~S., Cadwell~B.~J., Jacoby~B.~A., Wolszczan~A., Foster~R.~S.,
  Kramer~M., 1996, ApJ, 468, L55

\bibitem[Angerhofer \& Kundu~{1981}]{ak81}
Angerhofer~P.~E., Kundu~M.~R., 1981, AJ, 86, 1003

\bibitem[Camilo {\rm et~al.}~{2002}]{cmg+02}
Camilo~F., Manchester~R.~N., Gaensler~B.~M., Lorimer~D.~L., Sarkissian~J.,
  2002, ApJ, 567, L71

\bibitem[{Cordes} \& {Lazio}~{2002}]{cl02}
{Cordes}~J.~M., {Lazio}~T.~J.~W., 2002, ApJ, submitted ({astro-ph/0207156})

\bibitem[{F\"urst} \& Reich~{1986}]{fr86}
{F\"urst}~E., Reich~W., 1986, A\&A, 163, 185

\bibitem[{F\"urst} {\rm et~al.}~{1982}]{frb+82}
{F\"urst}~E., Reich~W., Beck~R., Hirth~W., Angerhofer~P.~E., 1982, A\&A, 115,
  428

\bibitem[Heger {\rm et~al.}~{2003}]{hwls03}
Heger~A., Woosley~S.~E., Langer~N., Spruit~H.~C., 2003, in Maeder~A.,
  Eenens~P., eds, Stellar Rotation, Proc.~of IAU Symposium S215.
\newblock PASP, San Francisco, in press (astro-ph/0301374)

\bibitem[Hobbs {\rm et~al.}~{2003}]{hlk+03}
Hobbs~G., Lyne~A.~G., Kramer~M., Martin~C.~E.~J.~C., 2003a, MNRAS, in prep.

\bibitem[Hobbs, Lyne \& Kramer~{2003}]{hlk03}
Hobbs~G., Lyne~A.~G., Kramer~M., 2003b, in M.~Bailes~D.~Nice~.~S.~T., ed, Radio
  Pulsars (ASP Conf.~Ser.), 
\newblock PASP, San Francisco, in press (astro-ph/0211001)

\bibitem[Kundu {\rm et~al.}~{1980}]{kafh80}
Kundu~M.~R., Angerhofer~P.~E., {F\"urst}~E., Hirth~W., 1980, A\&A, 92, 225

\bibitem[Lange {\rm et~al.}~{2001}]{lcw+01}
Lange~C., Camilo~F., Wex~N., Kramer~M., Backer~D., Lyne~A., Doroshenko~O.,
  2001, MNRAS, 326, 274


\bibitem[Lyne \& Manchester~{1988}]{lm88}
Lyne~A.~G., Manchester, R.~N., 1988, MNRAS, 234, 477

\bibitem[Lyne, Pritchard \& Smith~{1993}]{lps93}
Lyne~A.~G., Pritchard~R.~S., Smith~F.~G., 1993, MNRAS, 265, 1003

\bibitem[Lyne, Pritchard \& Shemar~{1995}]{lps95}
Lyne~A.~G., {Pritchard} R.~S., {Shemar} S.~L., 1995, JApA, 16, 179


\bibitem[McGowan {\rm et~al.}~{2003}]{mkz+03}
McGowan~K.~E., Kenea~J.~A., Zane~S., Cordova~F.~A., Cropper~M., Ho~C.,
  Sasseen~T., Vestrand~W.~T., 2003, ApJ, in press (astro-ph/0303380)

\bibitem[Migliazzo {\rm et~al.}~{2002}]{mgb+02}
Migliazzo~J.~M., Gaensler~B.~M., Backer~D.~C., Stappers~B.~W., van~der
  Swaluw~E., Strom~R.~G., 2002, ApJ, 567, L141

\bibitem[Morris {\rm et~al.}~{1981}]{mgs+81}
Morris~D., Graham~D.~A., Seiber~W., Bartel~N., Thomasson~P., 1981, A\&AS, 46,
  421

\bibitem[{Pavlov} {\rm et~al.}~{2001}]{pzs+01}
{Pavlov}~G.~G., {Zavlin}~V.~E., {Sanwal}~D., {Burwitz}~V., {Garmire}~G.~P.,
  2001, ApJ, 552, L129

\bibitem[{Pavlov} {\rm et~al.}~{2002}]{pzst02}
{Pavlov}~G.~G., {Zavlin}~V.~E., {Sanwal}~D., {Tr{\" u}mper}~J., 2002, ApJ, 569,  L95

\bibitem[Radhakrishnan \& Cooke~{1969}]{rc69a}
Radhakrishnan~V., Cooke~D.~J., 1969, Astrophys. Lett., 3, 225

\bibitem[{Romani} \& {Ng}~{2003}]{rn03}
{Romani}~R.~W., {Ng}~C.-Y., 2003, ApJ, 585, L41

\bibitem[Slane, Helfand \& Murray~2002]{shm02}
{Slane}~P.~O., {Helfand}~D.~J., {Murray}~S.~S., 2002, ApJ, 571, L45

\bibitem[Sofue, F\"urst \& Hirth~1980]{sfh80}
{Sofue}~Y., {F\"urst}~E., {Hirth}~W., 1980, PASJ, 32, 1

\bibitem[Stairs, Lyne \& Shemar 2000]{sls00}
{Stairs} I.~H., {Lyne} A.~G., {Shemar}, S.~L., 2000, Nature,
406, 484

\bibitem[Yakolev {\rm et~al.}~{2002}]{ykhg02}
Yakolev~D.~G., Kaminker~A.~D., Haensel~P., Gnedin~O.~Y., 2002, A\&A, 389, L24

\bibitem[{Zavlin} \& {Pavlov}~{2002}]{zp02}
{Zavlin}~V.~E., {Pavlov}~G.~G., 2002, in Neutron Stars, Pulsars, and Supernova
  Remnants, eds.~W.~Becker, H.~Lesch, J.~Tr\"umper, MPE-Report 278, p.~263

\end{thebibliography}

\clearpage

\begin{deluxetable}{ll}
\tablecaption{\label{tab:parms}Timing parameter of PSR J0538+2817}
\tablecolumns{2}
\tablewidth{0pc}
\tablehead{
\colhead{Parameter}   & \colhead{Value}\\}
\startdata
Ecliptic longitude, $\lambda$ (deg) &  85.232553(3)  \\
Ecliptic latitude, $\beta$ (deg) &      4.93582(4)   \\
R.A.\tablenotemark{a} (J2000) & 05 38 25.0623\\
DEC\tablenotemark{a} (J2000) & 28 17 09.1\\
Epoch (MJD) &  51086.0 \\
Spin frequency, $\nu$ (s$^{-1}$)   &   6.985276348019(5)    \\
First derivative, $\dot{\nu}$ ($10^{-15}$ s$^{-2}$) &  $-$179.04753(6) \\
Second derivative, $\ddot{\nu}$ ($10^{-24}$ s$^{-3}$) & $-$0.637(2) \\
Spin period, $P$ (ms) &  143.1582589118(1) \\
First derivative, $\dot{P}$ ($10^{-15}$) & 3.669452(1) \\
Dispersion Measure, DM (cm$^{-3}$ pc) & 39.814(6) \\
Proper motion, $\mu_\lambda$ (mas yr$^{-1}$) & $-$41(3) \\
Proper motion, $\mu_{\beta}$ (mas yr$^{-1}$) & 47(57) \\
Proper motion, composite\tablenotemark{b} (mas yr$^{-1}$) & $67^{-22}_{+48}$ \\
TOA Span (MJD) & 49453 -- 52712 \\
Number of TOAs & 249 \\
Timing RMS ($\mu$s) & 144.2 \\
\enddata
\tablecomments{The uncertainties in the last
quoted digits are given in parenthesis.}
\tablenotetext{a}{calculated from ecliptic coordinates}
\tablenotetext{b}{quoted value is median of asymmetric distribution}

\end{deluxetable}

\clearpage

\begin{figure}[ht]
\epsscale{0.8}
\plotone{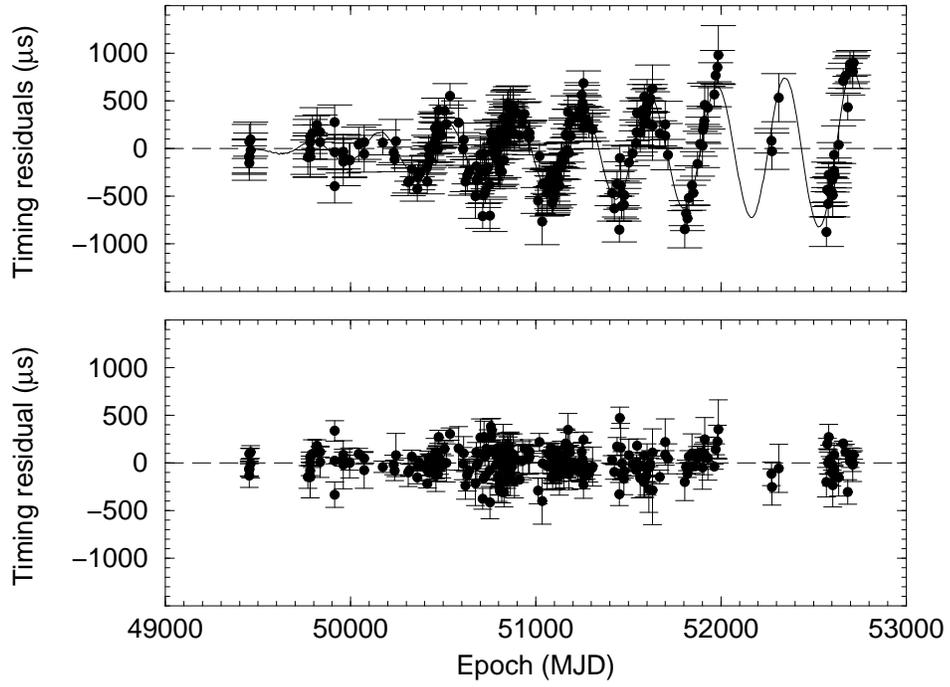}

\caption{ \label{fig:res}
Timing residuals obtained after applying a spin-down model
listed in Table 1 with a fit for proper motion (bottom)
and with a corresponding proper motion set to zero (top).
The solid line in the upper plot shows the expected 
behaviour for residuals in the latter case.}
\end{figure}

\clearpage

\begin{figure}[ht]
\epsscale{0.8}
\plotone{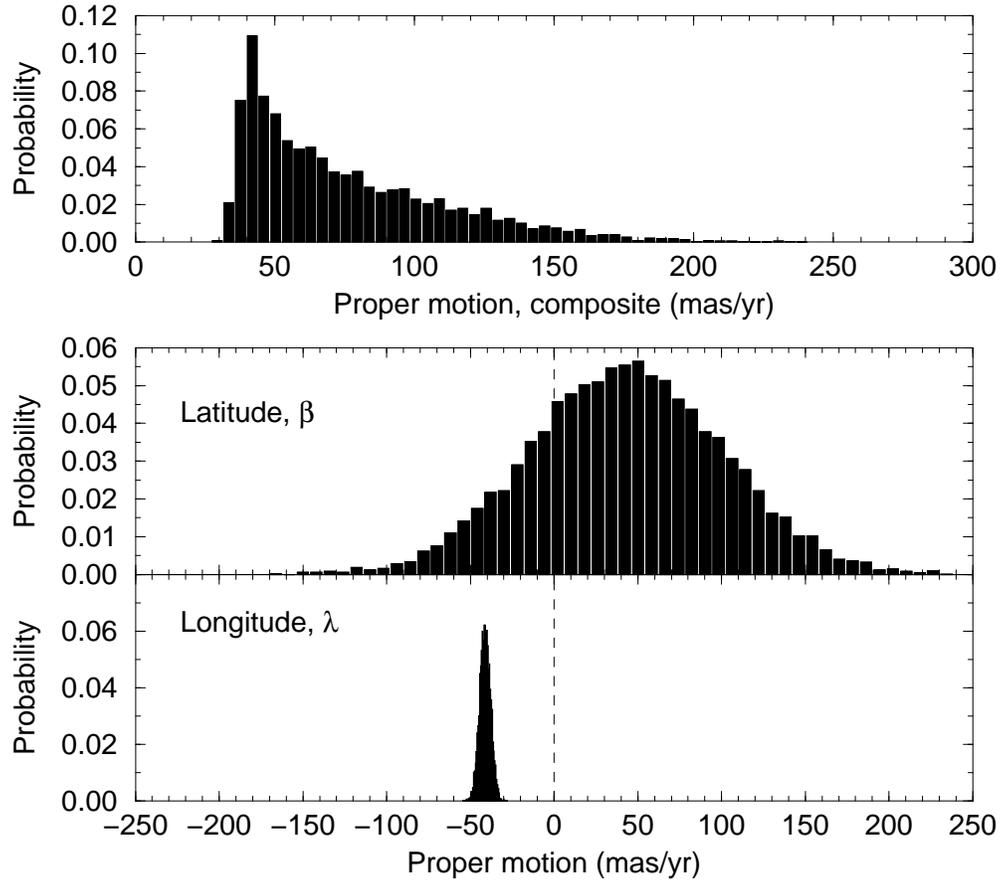}

\caption{ \label{fig:mc}
Results of Monte-Carlo simulations to determine
the uncertainties in proper motion measurements.}
\end{figure}

\clearpage

\begin{figure}[ht]
\epsscale{0.8}
\plotone{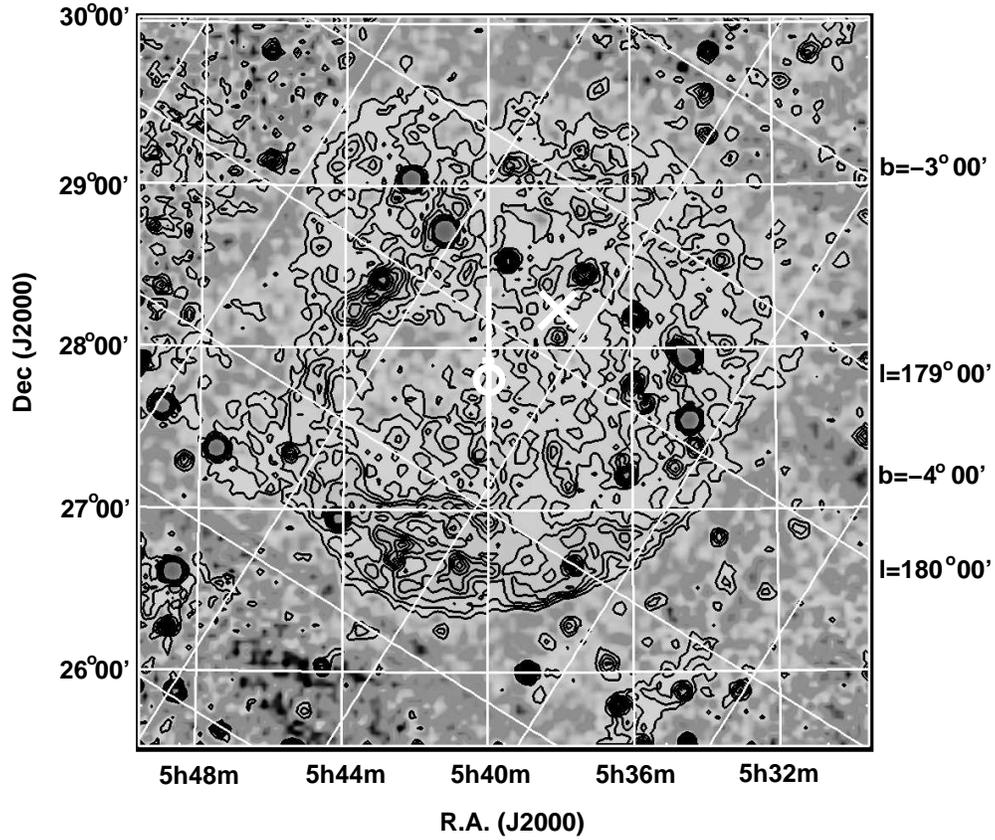}

\caption{
\label{fig:map}Radio image of S147 obtained at
2.7 GHz by F\"urst \& Reich~(1986) convolved to a 5' beam. Contours
are in steps of 25 mK $T_B$ beginning at $27.5$ mK $T_B$. The
right scale indicates Galactic coordinates.  The
determined center of the SNR is marked by a circle, the position of
PSR J0538+2817 by a 'X'. The position of the pulsar 30,000 years ago
(uncertainties are marked) agrees well with the centre of the SNR.}
\end{figure}

\end{document}